\documentclass[prl,twocolumn,showpacs,preprintnumbers,amsmath,amssymb]{revtex4}

\usepackage{graphicx}
\usepackage{dcolumn}
\usepackage{bm}
\newcommand{\bra}[1]{\langle#1|}

\newcommand{\modul}[1]{|#1|}
\newcommand{\scal}[2]{\langle#1|#2\rangle}\newcommand{\ket}[1]{|#1\rangle}

\begin{document}

\preprint{APS/123-QED}

\title{Single-Shot Generation and Detection of a Two-Photon Generalized Binomial State\\ in a Cavity}

\author{R. Lo Franco}
\email{lofranco@fisica.unipa.it}
\homepage{http://www.fisica.unipa.it/~lofranco}
\author{G. Compagno}
\author{A. Messina}
\author{A. Napoli}

\affiliation{%
Dipartimento di Scienze Fisiche ed Astronomiche, Universit\`{a} di
Palermo, via Archirafi 36, 90123 Palermo, Italy }

\date{\today}

\begin{abstract}
A ``quasi-deterministic'' scheme to generate a two-photon
generalized binomial state in a single-mode high-$Q$ cavity is
proposed. We also suggest a single-shot scheme to measure the
generated state based on a probe two-level atom that ``reads'' the
cavity field. The possibility of implementing the schemes is
discussed.
\end{abstract}


\maketitle

Generation of nonclassical states of the electromagnetic field holds
an important role in quantum optics both from the theoretical and
experimental point of view. In fact, these states may give
information about fundamentals of quantum theory and lead to
applications in quantum information processing \cite{zei,lof}. Due
to the experimental improvement of the quality factors of the
cavities, Rydberg atoms lifetimes and control of the atom-cavity
interactions \cite{har,rai,har1,har3}, cavity quantum
electrodynamics (CQED) is particularly indicated for quantum field
state engineering. In this context, several schemes have been
proposed to generate, for example, Fock states using the interaction
of consecutive atoms with a high-$Q$ cavity \cite{mey,har}. Recently
a two-photon Fock state was generated and probed \cite{bert}.

An important class of quantum non-classical states of the
electromagnetic field is constituted by the binomial states,
introduced by \citet{sto}, whose properties \cite{sto,vid,dat,vid1}
and interaction with atoms \cite{jos2} have been studied. These
states exhibit non-zero field expectation values, are characterized
by a finite maximum number of photons and interpolate between the
coherent state and the number state. They also have interesting
applications. For example, binomial states have been proposed as
reference field states in schemes to measure the canonical phase of
quantum electromagnetic fields \cite{peg3,peg4}. Generation of
entanglement between atoms and electromagnetic field was analyzed
when a binomial state interacts with a mixed two-qubit system
(two-level atoms) \cite{mah}. It was also recently shown that a
binomial state gives an interesting transient spectrum when it
constitutes the initial field state of a single-Cooper-pair box
\cite{mah1}. Thus, for its characteristic features and applications,
it appears of interest to develop implementable procedures for the
generation of binomial states.

A conditional scheme to generate binomial states in a cavity was
proposed \cite{mou}, in the CQED context, that exploits the quantum
field state engineering in a single-mode cavity introduced by
\citet{vog}. This scheme utilizes the resonant interaction of $N$
consecutive two-level atoms with the cavity initially prepared in
its vacuum state. The desired cavity field state is then obtained
through a total state reduction by a measurement on the atoms coming
out of the cavity. This scheme is conditional and results to have a
low efficiency for generating binomial states with a maximum number
of photons larger than one.

Here we propose an efficient, ``quasi-deterministic'' scheme for the
generation and detection of a two-photon generalized binomial state
in a single-mode high-$Q$ cavity. Moreover, we discuss its
implementation by considering the typical experimental errors
involved in CQED systems.

Our generation scheme exploits two consecutive two-level atoms
resonantly interacting, one by one, with the cavity initially
prepared in its vacuum state. The interaction of each two-level atom
with the single-mode cavity field is assumed to be well described by
the Jaynes--Cummings Hamiltonian
$H_{JC}=\hbar\omega\sigma_{z}/2+\hbar\omega a^{\dag}a+i\hbar
g(\sigma_{+}a-\sigma_{-}a^{\dag})$ \cite{jay}, where $\omega$ is the
cavity field mode, $g$ the atom-field coupling constant, $a$ and
$a^{\dag}$ the field annihilation and creation operators and
$\sigma_{z}=\ket{\uparrow}\bra{\uparrow}-\ket{\downarrow}\bra{\downarrow}$,
$\sigma_{+}=\ket{\uparrow}\bra{\downarrow}$,
$\sigma_{-}=\ket{\downarrow}\bra{\uparrow}$ the pseudo-spin atomic
operators, $\ket{\uparrow}$ and $\ket{\downarrow}$ being
respectively the excited and ground state of the two-level atom. It
is well known that the Hamiltonian $H_{JC}$ generates the
transitions \cite{mey,comp}
\begin{eqnarray}
\ket{\uparrow n}&\rightarrow&\cos(g\sqrt{n+1}t)\ket{\uparrow n}-\sin(g\sqrt{n+1}t)\ket{\downarrow n+1}\nonumber\\
\ket{\downarrow n}&\rightarrow&\cos(g\sqrt{n}t)\ket{\downarrow
n}+\sin(g\sqrt{n}t)\ket{\uparrow n-1},\label{evo}
\end{eqnarray}
where $\ket{\uparrow n}\equiv\ket{\uparrow}\ket{n}$,
$\ket{\downarrow n}\equiv\ket{\downarrow}\ket{n}$ and $a^\dag
a\ket{n}=n\ket{n}$.

It is useful, at this point, to recall that the single-mode
generalized binomial state (GBS) with a maximum number of photons
$N$, normalized and characterized by the probability of single
photon occurrence $0\leq p\leq1$ and the mean phase $\phi$ is
defined as \cite{sto,vid,peg1,peg2}
\begin{equation}
\ket{N,p,\phi}=\sum_{n=0}^N\left[{N\choose
n}p^{n}(1-p)^{N-n}\right]^{1/2}e^{in\phi}\ket{n}.\label{bin}
\end{equation}
This GBS is reduced to the vacuum state $\ket{0}$ when $p=0$ and to
the number state $\ket{N}$ when $p=1$. In the particular case where
$\phi=0$, the GBS of Eq.~(\ref{bin}) is named ``binomial state''
\cite{sto}. Here we concentrate on the 2GBS $\ket{2,p,\phi}$ that
can be explicitly written putting $N=2$ in Eq.~(\ref{bin})
\begin{equation}
\ket{2,p,\phi}=(1-p)\ket{0}+\sqrt{2p(1-p)}e^{i\phi}\ket{1}+pe^{i2\phi}\ket{2}.\label{bin2}
\end{equation}

\begin{figure}
\includegraphics[width=0.46\textwidth]{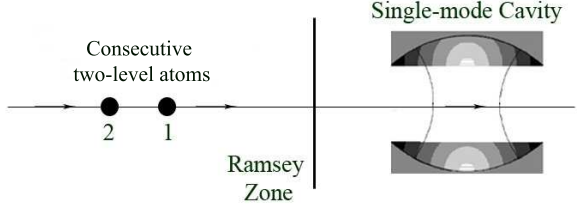}
\caption{Experimental setup for the generation of a two-photon
generalized binomial state (2GBS) in a cavity.}\label{figatomcav}
\end{figure}
A sketch of our generation procedure is represented in
Fig.\ref{figatomcav}. Before entering the cavity, the $k$-th
two-level atom ($k=1,2$) is injected into an appropriate Ramsey zone
where it is prepared in the superposition
\begin{equation}
\ket{\chi_k}=\sqrt{p}\ket{\uparrow}+e^{i\varphi_k}\sqrt{1-p}\ket{\downarrow}\quad(k=1,2)\label{chi}
\end{equation}
where $0\leq p\leq1$. The values of $p$ and $\varphi_k$ can be
arbitrarily fixed by adjusting the Ramsey zone settings, i.e. the
classical field amplitude and the atom-field interaction time. After
crossing the Ramsey zone, the first atom resonantly interacts for a
time $T_1$ with the cavity initially in the vacuum state $\ket{0}$.
Using Eq.~(\ref{evo}) together with Eq.~(\ref{chi}) and choosing an
interaction time $T_1=\pi/2g$, we obtain the factorized total
atom-cavity state
$\ket{\Psi_1(T_1)}=e^{i\varphi_1}\ket{p,\pi-\varphi_1}\ket{\downarrow}$
where the cavity field state
\begin{equation}
\ket{p,\pi-\varphi_1}=\sqrt{1-p}\ket{0}+e^{i(\pi-\varphi_1)}\sqrt{p}\ket{1},\label{bin1}
\end{equation}
is the GBS with $N=1$ and mean phase $\phi=\pi-\varphi_1$, also
called ``generalized Bernoulli state'' \cite{sto}. So, the field
state inside the cavity is generated, in principle, exactly in a
deterministic way. At this stage our scheme is identical to that one
proposed in Refs.~\cite{vog,mou}. It is worth to underline here that
this ($N=1$) GBS is measurable by a probe atom and this provides
also a method to test a Bell's inequality violation for an
entanglement of these states in two separate cavities \cite{lof}.

After the exit from the cavity of the first atom, i.e. after
preparing the cavity in the state $\ket{p,\pi-\varphi_1}$, a time
interval $\Delta t$ will elapse before the second atom enters the
cavity. During this time, the free field evolution induces a shift
of the mean phase of the cavity field state equal to $-\omega\Delta
t$. So, the second two-level atom finds the cavity field in the
state $\ket{p,\pi-\varphi'}$ with $\varphi'=\varphi_1+\omega\Delta
t$. This second atom, prepared by the Ramsey zone in the
superposition $\ket{\chi_2}$ of Eq.~(\ref{chi}) with
$\varphi_2=\varphi'$, interacts with the cavity for a time $T_2$.
Exploiting once again Eq.~(\ref{evo}) together with
Eqs.~(\ref{chi}), (\ref{bin1}) and with $\varphi'$ in place of
$\varphi_1$, at the end of the atom-cavity interaction the state of
the total system can be written, within a global phase factor, as
follows
\begin{eqnarray}
\ket{\Psi_2(T_2)}&=&\Big[(\cos gT_2-\sin
gT_2)\sqrt{p(1-p)}e^{i(\pi-\varphi')}\ket{0}\nonumber\\
&+&\cos(g\sqrt{2}T_2)pe^{2i(\pi-\varphi')}\ket{1}\Big]\ket{\uparrow}-\Big[(1-p)\ket{0}\nonumber\\
&+&(\sin gT_2+\cos
gT_2)\sqrt{p(1-p)}e^{i(\pi-\varphi')}\ket{1}\nonumber\\
&+&\sin(g\sqrt{2}T_2)pe^{2i(\pi-\varphi')}\ket{2}\Big]\ket{\downarrow}.\label{Psi2}
\end{eqnarray}
It is immediate to observe that, if the following conditions were
simultaneously satisfied
\begin{equation}
\sin(gT_2+\pi/4)=1,\quad\sin(g\sqrt{2}T_2)=1,\label{bincond}
\end{equation}
then the state $\ket{\Psi_2(T_2)}$ would be reduced to a factorized
state given by the product of the ground atomic state
$\ket{\downarrow}$ and the 2GBS $\ket{2,p,\pi-\varphi'}$ of
Eq.~(\ref{bin2}). Unfortunately the conditions of
Eq.~(\ref{bincond}) cannot be simultaneously satisfied. However, we
shall see that, by satisfying the first condition of
Eq.~(\ref{bincond}), the function $\sin(g\sqrt{2}T_2)$ takes a value
different from one for an amount $\delta$ smaller than the
uncertainty due to the typical experimental errors. In fact, the
first condition of Eq.~(\ref{bincond}) is satisfied for
$gT_2=\pi/4+2m_2\pi$, where $m_2$ is a non-negative integer. We now
look for suitable values of $m_2$ such that the second condition of
Eq.~(\ref{bincond}) is as near as possible to one. The typical
experimental conditions limit the interaction times $T$ in CQED
systems inside the range $10^{-1}\leq gT\leq10^2$ \cite{har}, and
these in turns confine the possible values of $m_2$ inside the
interval $0\leq m_2\leq16$. Among these possible values of $m_2$, we
find numerically that the best approximation of the second condition
of Eq.~(\ref{bincond}) occurs for $m_2=5$ which corresponds to
$T_2=41\pi/4g$. For this value of $T_2$ we have
$\sin(g\sqrt{2}T_2)=1-\delta$ where $\delta\sim10^{-4}$. On the
other hand, the deviation $\delta_{exp}$, induced by typical
experimental errors, may be estimated as
$\delta_{exp}\approx2(gT_2)^2(\Delta T_2/T_2)^2$ and, being $\Delta
T_2/T_2\sim10^{-2}$ \cite{har1,hag}, we have
$\delta_{exp}\approx10^{-1}$ hence $\delta\ll\delta_{exp}$. Summing
up, in correspondence to the interaction time $T_2=41\pi/4g$, the
cavity field state factor of $\ket{\downarrow}$ in Eq.~(\ref{Psi2})
is given by
\begin{equation}
\ket{\psi_{2,p,\pi-\varphi'}}\equiv
\frac{1}{\mathcal{N}_2}\sum_{n=0}^2c_n^{(2)}\left[p^{n}(1-p)^{2-n}\right]^{1/2}e^{in(\pi-\varphi')}\ket{n}
\label{psi2}
\end{equation}
where the coefficients are $c_0^{(2)}=1$, $c_1^{(2)}=\sqrt{2}$,
$c_2^{(2)}=1-\delta$ and ${\cal N}_2\approx(1-2\delta p^2)^{1/2}$ is
a normalization constant. In order to estimate how much the state
$\ket{\psi_{2,p,\pi-\varphi'}}$ is near to the ``target'' 2GBS
$\ket{2,p,\pi-\varphi'}$, we use the fidelity
$\mathcal{F}(p)=\modul{\scal{2,p,\pi-\varphi'}{\psi_{2,p,\pi-\varphi'}}}^2$.
This tends to one when $p\rightarrow0$, i.e. when the 2GBS is
reduced to the vacuum state $\ket{0}$. However, for $p\neq0$ the
fidelity is near to one and for $p=1/2$ we have
$\mathcal{F}(1/2)\approx1-1.6\times10^{-9}$. So,
$\ket{\psi_{2,p,\pi-\varphi'}}$ can be effectively identified with
the 2GBS $\ket{2,p,\pi-\varphi'}$ of Eq.~(\ref{bin2}). The
probability ${\cal P}_2$ to generate the cavity field state of
Eq.~(\ref{psi2}) is equal to the probability of finding the second
atom in the ground state $\ket{\downarrow}$ after coming out of the
cavity and it results to be, at the time $T_2=41\pi/4g$, ${\cal
P}_2=\modul{\bra{\downarrow}\scal{\psi_{2,p,\pi-\varphi'}}{\Psi_2(T_2)}}^2={\cal
N}_2^2$. Substituting the numerical value of $\mathcal{N}_2$ given
above, the probability to generate the 2GBS is
\begin{equation}
{\cal P}_2\approx1-2\times10^{-4}p^2,\label{pro2}
\end{equation}
that is much higher than the analogous generation probability of
previous conditional schemes ($\sim1/4$) \cite{vog,mou}. Moreover,
because of the high generation probability of the 2GBS of
Eq.~(\ref{pro2}) and being the typical atomic detection efficiency
less than one ($\sim70\%\div80\%$), the use of atomic detectors to
collapse the total state to the target cavity state cannot further
reduce the uncertainty on the generated cavity state itself. So,
within the experimental limits, in our generation scheme a final
atomic measurement is not required and it can be considered
non-conditional: for this reason we name it ``quasi-deterministic''.

\begin{figure}
\includegraphics[width=0.46\textwidth]{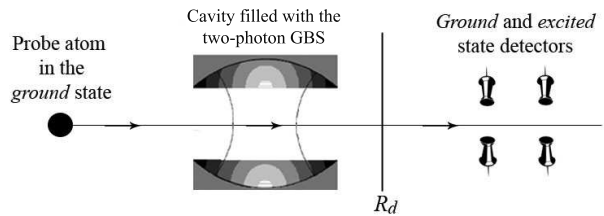}
\caption{\label{fig2}Experimental setup for detecting the 2GBS
inside the cavity. $R_d$ is the ``decoding'' Ramsey zone.}
\end{figure}
At this point, in order to probe that the cavity is effectively
filled with the 2GBS $\ket{2,p,\phi}$ of Eq.~(\ref{bin2}), we
describe the following single-shot measurement scheme, illustrated
in Fig.\ref{fig2}. Let us consider the cavity prepared in the 2GBS
$\ket{2,p,\phi}$ and a probe two-level atom prepared in the state
$\ket{\downarrow}$ that resonantly interacts with the cavity for a
time $T_P=41\pi/4g$. We have seen that, for such a time, both the
equalities of Eq.~(\ref{bincond}) can be retained satisfied within
the experimental errors. Thus, using the Jaynes--Cummings evolutions
reported in Eq.~(\ref{evo}) together with Eq.~(\ref{bin2}), we find
that, after the time $T_P$, the total state of the atom-cavity
system is transformed as
\begin{equation}
\ket{\downarrow}\ket{2,p,\phi}\stackrel{T_P}{\longrightarrow}\ket{p,\phi}(\sqrt{1-p}\ket{\downarrow}+\sqrt{p}\
e^{i\phi}\ket{\uparrow}),\label{probe1}
\end{equation}
where $\ket{p,\phi}$ is the GBS defined by Eq.~(\ref{bin}) for
$N=1$. After coming out of the cavity the atom crosses a
``decoding'' Ramsey zone $R_d$ set in such a way that it undergoes
the following transformations
\begin{eqnarray}
\ket{\uparrow}&\stackrel{R_d}{\longrightarrow}
&\sqrt{p}\ket{\uparrow}-e^{-i\phi}\sqrt{1-p}\ket{\downarrow}\nonumber\\
\ket{\downarrow}&\stackrel{R_d}{\longrightarrow}&
e^{i\phi}\sqrt{1-p}\ket{\uparrow}+\sqrt{p}\ket{\downarrow},\label{ramseq}
\end{eqnarray}
with the values of $p$ and $\phi$ coinciding with that ones defining
the 2GBS $\ket{2,p,\phi}$ to be measured. Thus, utilizing
Eqs.~(\ref{probe1}) and (\ref{ramseq}), we find that after the
Ramsey zone $R_d$ the atom-cavity system undergoes the evolution
\begin{equation}
\ket{\downarrow}\ket{2,p,\phi}\stackrel{T_P,R_d}{\longrightarrow}e^{i\phi}\ket{p,\phi}\ket{\uparrow}.\label{probe2}
\end{equation}
In this way, the measurement of the excited atomic state
$\ket{\uparrow}$ at the end of the sequence of Fig.\ref{fig2}
corresponds to the detection of the 2GBS $\ket{2,p,\phi}$ inside the
cavity.

We stress that, considering the orthogonality property of binomial
states \cite{lof}, the 2GBS $\ket{2,1-p,\pi+\phi}$, orthogonal to
the previously generated 2GBS $\ket{2,p,\phi}$, can also be obtained
by our generation scheme above with the changes
$p\rightarrow1-p,\varphi_k\rightarrow\pi+\varphi_k$ ($k=1,2$) in
Eq.~(\ref{chi}), which are achievable by appropriate adjustments of
the Ramsey zone settings. On the other hand, if a probe two-level
atom initially prepared in the ground state $\ket{\downarrow}$ finds
the 2GBS $\ket{2,1-p,\pi+\phi}$ inside the cavity and follows the
same measurement scheme as above, the atom-cavity evolution is
\begin{equation}
\ket{\downarrow}\ket{2,1-p,\pi+\phi}\stackrel{T_P,R_d}{\longrightarrow}\ket{1-p,\pi+\phi}\ket{\downarrow}\label{probe3}
\end{equation}
and the measurement of the ground atomic state $\ket{\downarrow}$ at
the end of the sequence of Fig.\ref{fig2} corresponds now to the
detection of the 2GBS $\ket{2,1-p,\pi+\phi}$ inside the cavity. We
also note that the results of Eqs.~(\ref{probe2}) and (\ref{probe3})
permit to distinguish, in a single-shot measurement, each of the two
orthogonal 2GBSs $\ket{2,p,\phi}$ and $\ket{2,1-p,\pi+\phi}$ inside
the cavity.

The orthogonal 2GBSs $\ket{2,p,\phi},\ket{2,1-p,\pi+\phi}$ represent
two vectors of an orthonormal basis in a 3-dimensional Fock--Hilbert
space. The third vector of the basis can be readily obtained by
setting the orthogonality and normalization conditions and it
results to be
\begin{eqnarray}
\ket{\Gamma(2,p,\phi)}&=&\sqrt{2p(1-p)}\ket{0}+(2p-1)e^{i\phi}\ket{1}\nonumber\\
&-&\sqrt{2p(1-p)}e^{i2\phi}\ket{2}.\label{Gamma}
\end{eqnarray}
This state can be generated in a conditional way \cite{lof1} by
using the resonant interaction of an opportunely prepared two-level
atom with a cavity filled with a GBS with $N=1$. Moreover, utilizing
the Holstein--Primakoff operators \cite{fu} for the case $N=2$,
$J_2^+=\sqrt{2-\hat{n}}a$, $J_2^-=a^\dag\sqrt{2-\hat{n}}$ and
$J_2^0=1-\hat{n}$ with $\hat{n}=a^\dag a$, we shall show that these
three basis states, $\ket{2,p,\phi}$,$\ket{\Gamma(2,p,\phi)}$ and
$\ket{2,1-p,\pi+\phi}$ are eigenvectors of the pseudo angular
momentum operator $J_3$ defined as
\begin{equation}
J_3=\sqrt{p(1-p)}\left(e^{-i\phi}J_2^{+}+e^{i\phi}J_2^-\right)-(2p-1)J_2^0.\label{J3}
\end{equation}
In fact, using Eq.~(\ref{J3}) and the explicit forms of the basis
vectors given in Eqs.~(\ref{bin2}), (\ref{Gamma}), it is
straightforward to prove the following eigenvalues equations
\begin{eqnarray}
J_3\ket{2,p,\phi}=\ket{2,p,\phi};\quad
J_3\ket{\Gamma(2,p,\phi)}=0\nonumber\\
J_3\ket{2,1-p,\pi+\phi}=-\ket{2,1-p,\pi+\phi}.
\end{eqnarray}
So, using the usual notation $\ket{l,m}$ for the eigenvectors of
angular momentum $l$, we can do the identifications
$\ket{2,p,\phi}\equiv\ket{1,1}$,
$\ket{\Gamma(2,p,\phi)}\equiv\ket{1,0}$,
$\ket{2,1-p,\pi+\phi}\equiv\ket{1,-1}$, and we can describe the
3-dimensional Fock--Hilbert subspace as a subspace of angular
momentum $l=1$ spanned by the basis
$\mathcal{B}=\{\ket{1,1},\ket{1,0},\ket{1,-1}\}$. In this context,
we say that our measurement scheme of orthogonal 2GBSs, illustrated
in Fig.\ref{fig2} with the results given by Eqs.~(\ref{probe2}),
(\ref{probe3}), constitutes a method to measure the eigenvalues
$\pm1$ of the field operator $J_3$ of Eq.~(\ref{J3}).

We now discuss the possible implementation of our
quasi-deterministic schemes. In our generation scheme the
atom-cavity interaction times are different for each atom,
respectively $T_1,T_2$. These can be obtained either by selecting
opportune different velocities for each atom or by selecting the
same velocity for the two atoms and applying an electric field
inside the cavity in order to Stark shift each atom out of resonance
so to obtain the desired resonant interaction time \cite{davi1}. The
appropriate atomic velocity may be selected by laser induced atomic
pumping \cite{har1}. The experimental uncertainties of the selected
velocity $\Delta v$ and interaction time $\Delta T$ are such that
$\Delta T/T\approx\Delta v/v$. In current laboratory experiments it
is possible to select a given atomic velocity such that $\Delta
v/v\sim10^{-2}$ or less \cite{har1,hag}. In our generation and
measurement schemes, we have also ignored the atomic or photon decay
during the atom-cavity interactions. This assumption is valid if
$\tau_{at},\tau_{cav}>T$, where $\tau_{at},\tau_{cav}$ are the
atomic and photon mean lifetimes respectively and $T$ is the
interaction time. For Rydberg atomic levels and microwave
superconducting cavities with quality factor $Q\sim10^8\div10^{10}$
the required condition on the mean lifetimes can be satisfied,
because $\tau_{at}\sim10^{-5}\div10^{-2}\textrm{s}$,
$\tau_{cav}\sim10^{-4}\div10^{-1}\textrm{s}$ and
$T\sim10^{-5}\div10^{-4}\textrm{s}$ \cite{har}. Moreover, the
typical mean lifetimes of the Rydberg atomic levels $\tau_{at}$ must
be such that the atoms do not decay during the entire sequence of
the schemes \cite{har,har1}.

In conclusion, we have shown that it is possible to generate, within
the experimental errors in a non-conditional way, a two-photon
generalized binomial state (2GBS) inside a single-mode cavity by
using two consecutive two-level atoms interacting with the cavity
each for a given time. Moreover, the presence inside the cavity of
the 2GBS can be verified by a single-shot measurement scheme
utilizing a probe two-level atom, prepared in its ground state, that
resonantly interacts with the cavity for a given time and ``reads''
the cavity field state. The information acquired by the probe atom
is then ``decoded'' by a suitable Ramsey zone and finally ``read''
by measuring the internal atomic state. The results of this work
therefore open the way to generation and detection schemes of
superpositions of two orthogonal 2GBSs in a cavity (``binomial
Schr\"{o}dinger cat'') or entangled 2GBSs in separate cavities
\cite{lof1}, which can be useful both for investigations of the
foundations of quantum theory and for applications in quantum
information processing. An extension of the scheme proposed here is
possible for an efficient generation of GBSs with $N>2$ and it will
be treated somewhere else. At this time, the experimental
developments seem to be rather promising on the possibility of
implementing our schemes.


\end{document}